\documentclass[prl,twocolumn,superscriptaddress,showpacs,nobalancelastpage]{revtex4}

\usepackage[pdftex]{graphicx}
\usepackage{float}

\begin{document}

\title{Using a two-electron spin qubit to detect electrons flying above the Fermi sea}

\author{Romain Thalineau}
\affiliation{Univ. Grenoble Alpes, Inst NEEL, F-38042 Grenoble, France}
\affiliation{CNRS, Inst NEEL, F-38042 Grenoble, France}

\author{Andreas D. Wieck}
\affiliation{Lehrstuhl f\"ur Angewandte Festk\"orperphysik, Ruhr-Universit\"at Bochum, Universit\"atsstrasse 150, 44780 Bochum, Germany.}

\author{Christopher B\"auerle}
\affiliation{Univ. Grenoble Alpes, Inst NEEL, F-38042 Grenoble, France}
\affiliation{CNRS, Inst NEEL, F-38042 Grenoble, France}

\author{Tristan Meunier}
\affiliation{Univ. Grenoble Alpes, Inst NEEL, F-38042 Grenoble, France}
\affiliation{CNRS, Inst NEEL, F-38042 Grenoble, France}

\date{\today}


\begin{abstract}

We investigate experimentally the capacitive coupling between a two-electron spin qubit and flying electrons propagating in quantum Hall edge channels. We demonstrate that the qubit is an ultrasensitive and fast charge detector with the potential to allow single shot detection of a single flying electron. This work opens the route towards quantum electron optics at the single electron level above the Fermi sea.

\end{abstract}

\maketitle

During the last two decades, an important effort has been devoted to the control of nanocircuits at the single electron level. It is now possible to isolate a single electron in a trap \cite{tarucha}, to displace it deterministically \cite{hermelin,mcneil, thalineau} and to manipulate it quantum mechanically \cite{koppens,petta,hayashi,foletti,brunner}. An important requirement for these achievements is to have sensitive enough detectors to probe the presence of a single electron. Up to now, single electron detection is only possible while the electron is trapped within a quantum dot. For example, in an AlGaAs heterostructure two dimensional electron gas (2DEG), it is indeed possible to freeze for a sufficiently long time the electron in the quantum dot and detect it with a conventional on-chip electrometer \cite{field}. For quantum experiments using flying electrons \cite{ji,roulleau,michi,bocquillon,glattli}, this task is much more difficult: indeed the time of interaction between the detector and the flying electron usually does not exceed 1 ns and is fixed by the speed of the electron, the size of the on-chip electrometer and the width of the electronic wavepacket. This time is 2 orders of magnitude faster than the time needed to detect a single electron with the best on-chip charge detector demonstrated so far in a 2DEG \cite{barthel}.  This represents an important limitation for the investigation of quantum correlations in experiments with flying electrons. 

\begin{figure}[h!]
\includegraphics[width=3.4in]{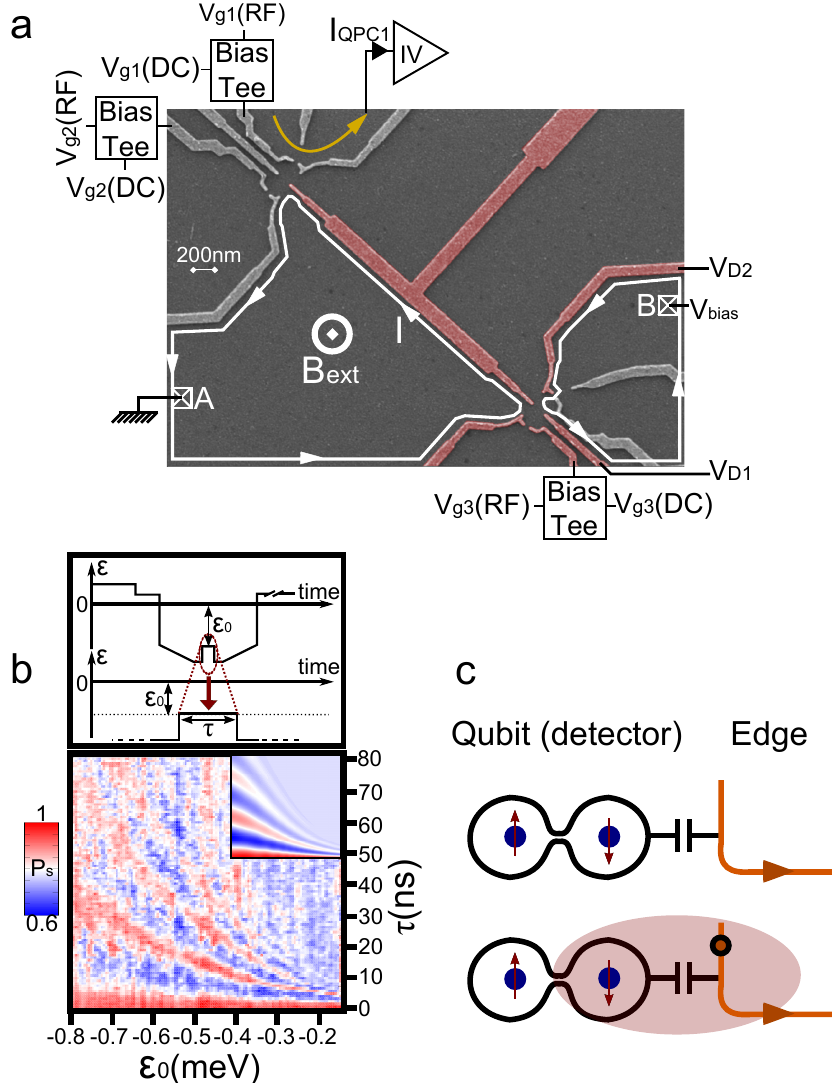}
\caption{\textbf{Experimental device and principle of the experiment.} \textbf{(a)} Scanning Electron Microscope image of the device and schematic of the experimental setup. Two multigate systems are separated by a 3 $\mu$m-long gate. In the upper left corner, the $S-T_0$ qubit detector made out of a double dot system that can be brought into the few-electron regime (see supplemental materials). It is probed with an on-chip electrometer with a net current represented with an orange arrow. In the bottom right corner, the multigate structure can either be used as a quantum point contact (QPC) or as a quantum dot (the gates polarized are represented in color, see supplemental materials). The two systems are interconnected via chiral edge channels (in white on the figure). Only the outer edge state is represented. \textbf{(b)} Coherent oscillations between antiparallel spin states (Singlet detected probability $P_s$ versus pulse interaction time $\tau$) for different $\epsilon_0$(see supplemental materials). The pulse sequence applied to $\epsilon$ is presented on the top of the figure. The inset presents the result of a fit procedure where $J$ depends exponentially on $\epsilon$ (see text). \textbf{(c)} Principle of the experiment: when an electron is passing by the $S-T_0$ qubit detector, it changes the chemical potential of the closest qubit dot to the edge state by capacitive coupling. It results in a change of the relative detuning between the qubit dots and therefore $J$.}
\label{Fig1}
\end{figure}

To overcome this limitation, quantum systems have been identified as extremely sensitive systems to external perturbations and potentially good detectors \cite{gleyzes}. They have been used for example to detect a single phonon excitation of a nanomechanical system \cite{oconnel}. In a 2DEG, two quantum systems have been recently proposed to detect a propagating electron: a double quantum dot charge qubit and a Mach-Zender interferometer \cite{neder}.

Here, we propose and demonstrate experimentally the potential of a two-electron spin qubit as an ultrasensitive charge detector by coupling it to individual electrons, propagating in the edge states of the Quantum Hall regime. The Singlet-Triplet ($S-T_0$) qubit in a double dot with two electrons is a quantum system where the two-electron antiparallel spin states can be manipulated on fast timescales by changing the energy detuning $\epsilon$ between the two dots. Moreover, the charge sensitivity of the qubit is tunable with $\epsilon$ and can be completely eliminated \cite{petta}. In this charge insensitive configuration, the quantum information stored in the $S-T_0$ qubit can be preserved for a time longer than a few hundreds of microseconds \cite{bluhm}. These properties allow interacting strongly for a very short time with a single electron, storing the resulting change in the population of the two-level system for a time sufficiently long and read-out the state of the qubit in a single shot with fast charge detection. In this way, single shot read out of flying electrons passing by the $S-T_0$ qubit could be performed.

We study a sample made of two Ti-Au multigate systems defined on top of a AlGaAs-GaAs heterostructure 2DEG separated by a 3 $\mu m$ long gate (see Fig. \ref{Fig1}(a), details on the sample can be found in the Supplemental materials). By applying a perpendicular magnetic field of 0.73 T, the electronic system enters the quantum Hall regime with a filling factor $\nu \sim 16$ and edge channels run along the gates. The gates on the upper part of the sample can be polarized with negative voltages in order to define the $S-T_0$ qubit (see Fig \ref{Fig1}(b) and supplemental materials). The electric field generated by the single electron propagating in the edge state will slightly modify the electrostatic environment of the double dot (see Fig. \ref{Fig1}(c)). It will therefore change the qubit energy splitting during the time of interaction and the frequency $J$ of the coherent oscillations between the antiparallel spin states. This results in a population change of the two-level system that can be stored for several tens of microseconds and then be measured by spin-to-charge conversion. In our experiment, we use two different ways to inject electrons into the edge state: either by a QPC or a fast tunable quantum dot (see Fig. \ref{Fig1}(a) and supplemental materials). 

The sensitivity of the $S-T_0$ detector to charge fluctuations can be evaluated from the $\epsilon$-dependence of the coherent oscillations presented in Fig. \ref{Fig1}(b). With a fit valid for $\epsilon<-0.3$ meV, we obtain $J(\epsilon)=J_0 + J_1 exp(-\epsilon/b)$ with $J_0=20$ neV, $J_1=2.5$ $\mu$eV and $b=200$ $\mu$eV. For a fixed detuning variation $\Delta \epsilon$, the relative change in $J$ is independent of $\epsilon$ and equal to $|(\Delta J/ \Delta \epsilon)/J|=0.005$ $\mu$eV$^{-1}$. Moreover the typical detector response timescale is set by the dynamics of the $S-T_0$ qubit. Therefore the more rapid the coherent oscillations are, the faster the detector responds. The most sensitive working point of the detector is then where the slope of the coherent exchange oscillations as a function of the interaction time is maximum. Knowing the shift detuning $\Delta \epsilon_e$ resulting from the coupling between a single flying electron and the $S-T_0$ detector permits evaluating the expected detector response for one electron.  

\begin{figure}
\includegraphics[width=3.4in]{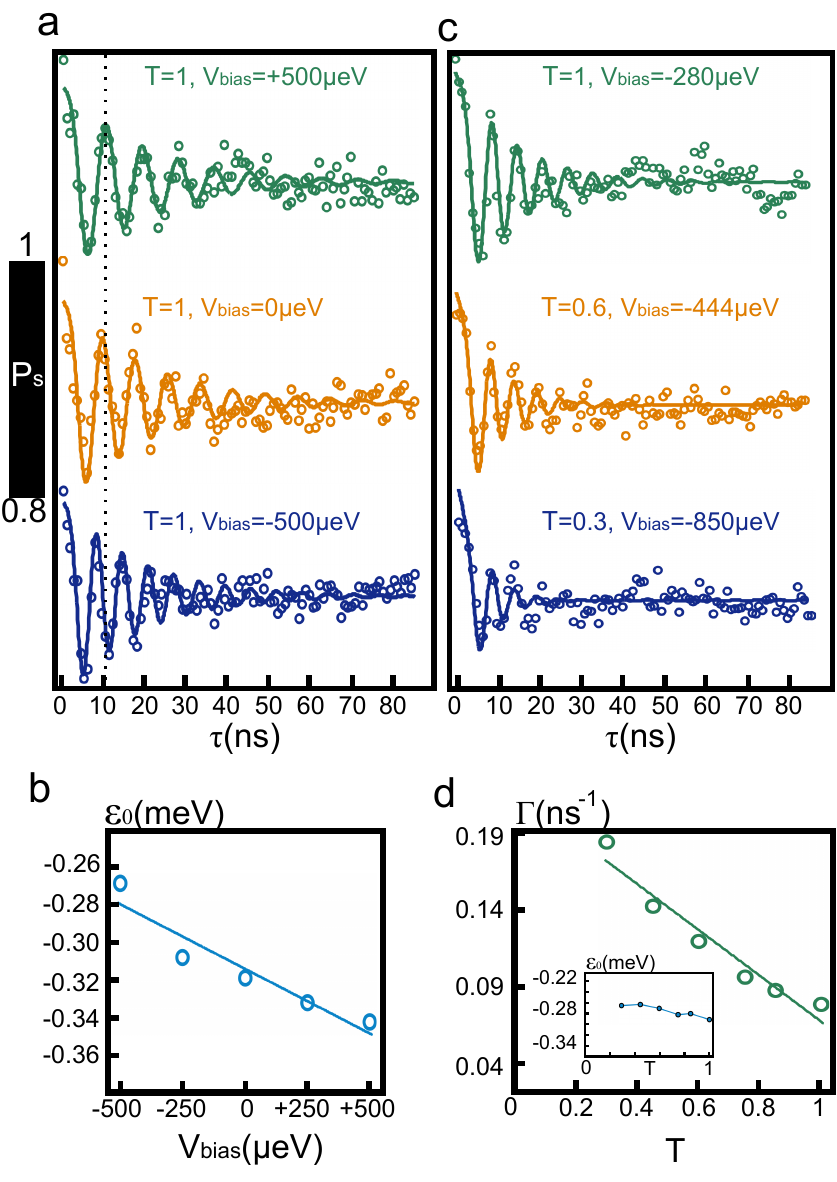}
\caption{\textbf{Electron injection in the edge state via a QPC.} \textbf{(a)} Coherent oscillations as a function of the bias applied on contact B at $T=1$. \textbf{(b)} $\epsilon_0$ as a function of the chemical potential of the edge state. \textbf{(c)} Coherent oscillations as a function of the transmission $T$ of the QPC by keeping the current $I$ fixed at 20nA. The solid lines correspond to numerical fit of the Schr\"odinger equation, governing the dynamics of the antiparallel spin states (see supplementary material). It allows for the extraction of the effective detuning pulse position $\epsilon_0$ and the decoherence time $\Gamma^{-1}$. \textbf{(d)} $\Gamma$ as a function of $T$ with $I$ fixed at 20 nA. The solid line is a linear fit with the slope of $0.14$ ns$^{-1}$. Inset: $\epsilon_0$ as a function of $T$ with $I$ fixed at 20 nA.}
\label{Fig2}
\end{figure}

To demonstrate experimentally the potential of such a detection method and evaluate the strength of $\Delta \epsilon_e$, a flow of electrons is injected into the edge state closest to the gates in a controllable manner by a QPC (see supplemental materials). Within the Landauer-B\"uttiker formalism \cite{buttiker}, the potential of the last edge state is fixed by the bias potential $V_{bias}$ applied to the contact B. When $V_{bias}$ is changed, we observe a variation of $J$ which results in faster (slower) oscillations when a negative (positive) bias is applied (see Fig. \ref{Fig2}(a) and \ref{Fig2}(b)). Indeed, the electric field induced by the closest edge state potential adds to the one generated by the gates and results in a change of $\epsilon$. In this low-exchange-coupling regime, a $\mu$eV shift in $\epsilon$ corresponds to a nA current flowing in the edge states ($\Delta I/ \Delta \epsilon=$1 nA/$\mu$eV), meaning that roughly an electron passes by every 0.2 ns. Such a detuning shift $\Delta \epsilon_e$ caused by the passage of a single electron corresponds to a change of the detector population of $0.01\%$.

By changing the transparency $T$ of the injector QPC barrier, we observe that the detector is sensitive not only to the average current in the channel but also to the current shot noise and therefore to electron granularity. For this purpose, we vary both $T$ and $V_{bias}$ such that the current $I$ in the edge state is kept constant (as shown in Fig. \ref{Fig2}(c)). As expected, $J$ remains nearly unchanged as we change the working point ($T$, $V_{bias}$) (see inset Fig. \ref{Fig2}(d)): the average number of electrons per second seen by the detector does not change. In addition, we observe a linear decrease of the decoherence rate $\Gamma$ (see Fig. \ref{Fig2}(d)). In the case where the coupling between the detector and the electrons is small in comparison with the typical time scale of the detector, the Gaussian approximation is applicable and the Bloch-Redfield theory predicts that the decoherence rate is directly proportional to the spectral density of current noise $S_I$ \cite{ithier} and equal to $\pi .S_I.(dJ/\hbar dI)^2$. In the situation of high bias relative to the probed frequency, the noise is expected to be frequency independent and the shot noise formula applies: $S_I(\omega)=(e^3 V/h) \times T(1-T)$. In keeping the current $V_{bias}.T$ constant, $S_I$ is linear in $1-T$ as the observed linear dependence of decoherence rate with $T$. The coupling $dJ/dI$ extracted from the fit of the decoherence rate (see Fig. \ref{Fig2}(d)) is an order of magnitude bigger than the one extracted by the exponentional fit of $J(\epsilon)$ (see Fig. \ref{Fig1}(c)) and the calibrated $dI/d\epsilon$. Nevertheless, the data presented in Fig. \ref{Fig2}(b) are taken at a position closer to the crossing between the (1,1) and (0,2) charge states and deviations from the exponential fit of the exchange coupling $J$ have been observed and result in higher sensitivity of the $S-T_0$ qubit to noise detuning \cite{dial}. Therefore, the behaviour of the coherent oscillations is in good agreement with the current shot noise theory of a single transmitted channel through a QPC \cite{martin} which allows us to assert that our detector is sensitive to individual flying electrons.

\begin{figure}
\includegraphics[width=3.4in]{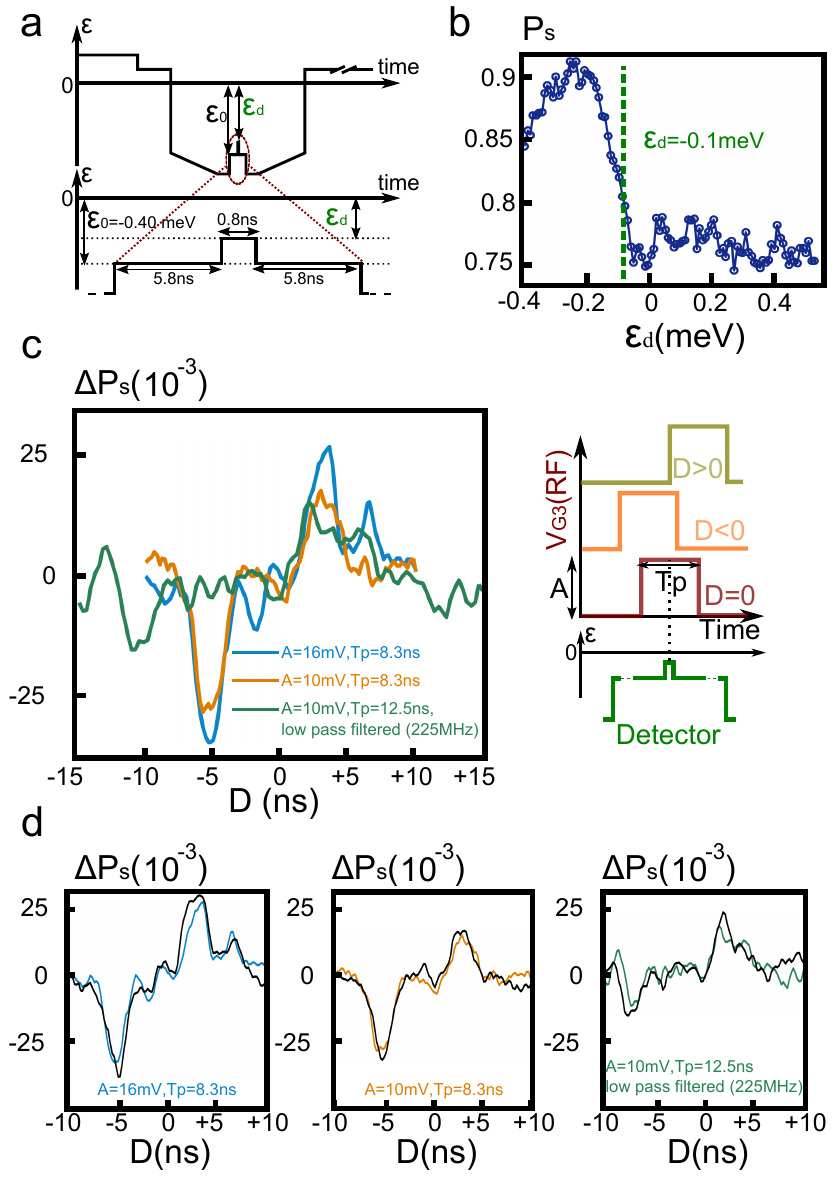}
\caption{\textbf{Nanosecond excitation and detection of the edge potential change. (a)} Detuning pulse sequence applied on the $S-T_0$ qubit to produce the nanosecond detection pulse. \textbf{(b)} Evolution of the population as a function of the nanosecond pulse amplitude $\epsilon_d$. An additional oscillation of $\pi$ is then observed. Then the coherent oscillations stop most likely because decoherence kills the phase coherent evolution. The most sensitive working point is set at $\epsilon_d = - 0.10$ meV. At this position, a positive (negative) local chemical potential variation of the edge channel induces an increase (decrease) of the singlet probability. \textbf{(c)} Left : Detector response, counted in population change $\Delta P_s$ with respect to the position where no excitation pulse is applied, as a function of the delay D for different pulse rise time, duration $T_p$ and amplitude $A$. Right : Schematics in the time domain of the pulse sequence applied to the detector and to the fast gate $V_{g3}$. The delay D is counted with respect to the center of the nanosecond detection pulse. \textbf{(d)} Comparison of the detector response when the source quantum dot is pulsed within a Coulomb Blockade region (colored lines) or between two different charge state regions (black lines) for different $A$, $T_p$ and pulse rise time.} 
\label{Fig3}
\end{figure}

The time-response of the detector is set by the quantum dynamics of the $S-T_0$ qubit and therefore nanosecond time-resolved detection is possible. To demonstrate this experimentally, we need to excite the edge state and arm the qubit-detector at nanosecond timescales. For this purpose, we formed a quantum dot in the lower right part of the sample to inject an electron by triggering $V_{g3}$ with a fast emission voltage pulse \cite{feve}. The width of the electron wave packet is set by the tunnel barrier separating the quantum dot from the edge state and is estimated to be between 0.1-1 ns (see supplemental material). In this way, we change the edge potential at nanosecond timescales. To arm the detector only for a nanosecond, we only pulse the detuning for a nanosecond (see Fig. \ref{Fig3}(a)). In this case, most of the evolution is happening when the pulse is close to its maximum value $\epsilon_d$ and therefore we estimate the relevant interaction time close to 0.5 ns. The resulting population evolution at fixed excitation time as a function of $\epsilon_d$ is presented in Fig. \ref{Fig3}(b). As expected, we reproduce a part of the coherent evolution. From $\Delta P_s/ \Delta \epsilon$ at $\epsilon_d = -0.1$ meV and the coupling between the detector and a single flying electron, we extract the expected change in the detector population $\Delta P_s$ induced by the passage of a single electron to about $0.2\%$. 

In Fig. \ref{Fig3}(c), the evolution of the detector signal or the probability to detect the singlet state is plotted as a function of the delay between the emission and detection pulses with a step size of 0.2 ns. The two pulses are synchronized using the same arbitrary waveform generator. In this curve, one observes two large peaks, one positive for positive delay and one negative for negative delay. They correspond to the falling and rising edges of the emission pulse and are consistent with the emission of an electron and a lack of electron happening at this position (see Fig. \ref{Fig3}(c)). Nevertheless, the amplitude of the peaks is larger than the estimated $\Delta P_s$ for a single flying electron. Moreover, when the emission pulse amplitude or rise time are reduced, this peak amplitude is diminishing. They are more likely the results of a local change in the electron density due to direct capacitive coupling between the pulsing gate $V_{g3}$ called edge magneto-plasmons EMP \cite{ashoori}. The excess or the lack of electrons is then provided by the coupling to the bulk at this low magnetic field regime. According to the extracted detector sensitivity, each pulse contains between 5 and 10 electrons.

The protocol described above is reproduced by pulsing the emitted quantum dot between two distinct charge state configurations or within a Coulomb blockade region. In this case, an electron or a lack of an electron is emitted in addition to the EMP only in the first case. Fig. \ref{Fig3}(d) presents the comparison between the resulting detection curves. For each curve, one notices a small increase of the peak amplitude that we assign to the emitted electron or lack of an electron. It indeed corresponds to the amplitude expected from the detector-electron coupling. The signal is however small compared to the one induced by the EMP and a clear demonstration will only be possible by eliminating the EMP.

In conclusion, we capacitively coupled a $S-T_0$ qubit to electrons propagating in a quantum Hall edge state. When a stream of electrons is emitted with a QPC, we demonstrate that the qubit detector is not only sensitive to the average current but also to the current shot noise. In pulsed experiments where a fast tunable quantum dot is used as an emitter, we demonstrated that the two-electron spin qubit can be considered as a sufficiently fast detector with a high sensitivity that allows to detect few propagating electrons. Moreover, we observe that the single electron is accompanied by a cloud of EMP excited during the emission pulse. For future experiments, it would be important to minimize as much as possible these extra excitations.  

In order to use such a detection method to detect quantum correlations in quantum optics experiments with flying electrons \cite{ji,roulleau,bocquillon}, one needs first to use lower density samples in order to work at lower filling factor and below 1 T where the qubit is working. Second, the single shot regime of detection needs to be achieved. This last requirement is within experimental reach. Indeed, the gate geometry could be optimized to increase the coupling between the edge channel and the $S-T_0$ qubit and, more importantly, more complex qubit operations and faster detection schemes could be used in order to increase the fidelity of the qubit operation as demonstrated in \cite{shulman}.

We acknowledge technical support from the "Poles" of the Institut N\'eel as well as Pierre Perrier. We acknowledge the help of S. Hermelin in the device fabrication. Devices were fabricated at "Plateforme Technologique Amont" of Grenoble, with the financial support of the "Nanosciences aux limites de la Nano\'electronique" Foundation and CNRS Renatech network. A.D.W. acknowledges expert help from PD Dr. Dirk Reuter and support of the DFG SPP1285 and the BMBF QuaHLRep 01BQ1035. R. T. and T.M. acknowledges financial support from ERC "QSPINMOTION".

\end{document}


\title{Supplemental material : Using a two-electron spin qubit to detect electrons flying above the Fermi sea}

\author{Romain Thalineau}
\affiliation{Univ. Grenoble Alpes, Inst NEEL, F-38042 Grenoble, France}
\affiliation{CNRS, Inst NEEL, F-38042 Grenoble, France}

\author{Andreas D. Wieck}
\affiliation{Lehrstuhl f\"ur Angewandte Festk\"orperphysik, Ruhr-Universit\"at Bochum, Universit\"atsstrasse 150, 44780 Bochum, Germany.}

\author{Christopher B\"auerle}
\affiliation{Univ. Grenoble Alpes, Inst NEEL, F-38042 Grenoble, France}
\affiliation{CNRS, Inst NEEL, F-38042 Grenoble, France}

\author{Tristan Meunier}
\affiliation{Univ. Grenoble Alpes, Inst NEEL, F-38042 Grenoble, France}
\affiliation{CNRS, Inst NEEL, F-38042 Grenoble, France}

\date{\today}


\maketitle

\textbf{Sample and experimental set-up}

The device is defined by Schottky gates in a n-AlGaAs/GaAs 2DEG based heterostructure (2DEG:  mobility $10^6$ $cm^2/Vs$ , density $2.7 \times 10^{11}$ $cm^{-2}$, depth = 90 nm) using standard split-gate techniques. The charge configuration of the double dot system (see Fig. 1 of main text) is measured via the conductance of a QPC by biasing it with a DC voltage of 500 $\mu$V and the current is measured with an IV-converter with a bandwidth of 1.4 kHz. The voltage potential of the three gates Vg1, Vg2 and Vg3 can be varied at nanosecond timescale using a Tektronix Arbitrary Waveform Generator 5014C with 0.9 ns rise time. The sample is anchored at the mixing chamber of a dilution refrigerator with an electron temperature of T=150 mK. The magnetic field applied to obtain a filling factor of 16 is 0.73~T.

\textbf{Coherent Exchange oscillation sequence}
\begin{figure}
\includegraphics[width=3.4in]{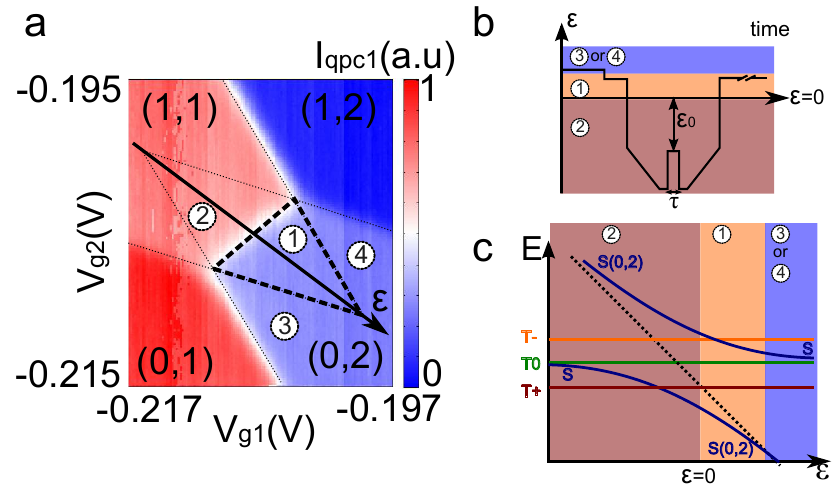}
\caption{\textbf{Stability diagram of the double dot system and coherent oscillation pulsing sequence (a)} Stability diagram of the double dot system close to the (1,1) - (0,2) crossing. The arrow represents the detuning axis. The numbers represent the different position of the system during the pulse sequence. \textbf{(b)} Corresponding detuning pulse sequence. \textbf{(c)} Energy diagram as a function of the detuning $\epsilon$ of the two-electron states in a double quantum dot. $S$ and ($T_+, T_0, T_-$) correspond respectively to the singlet and the triplet states when the two electrons are separated in the two dots, S(0,2) is the singlet state when they are both in the right dot.}
\label{FigS2}
\end{figure}

With two electrons in a double dot system, the control of the detuning parameter $\epsilon$ between the two dots allows manipulating quantum mechanically antiparallel spin states. The stability diagram and pulse scheme used is presented in Fig. \ref{FigS2}. The procedure is made of four stages and has been pioneered by Petta et al \cite{petta}. The system is initialized in the singlet state of the (0,2) charge configuration via spin relaxation by waiting 200 ns (in position 3 or 4). It is then pulsed in the measurement position 1 for 200 ns. Subsequently, it is brought in the (1,1) charge configuration region (position 2) adiabatically with respect to the tunnel coupling. The pulse is decomposed into two parts: the first one is fast in order to cross the $S-T_+$ crossing non-adiabatically, the second part of the pulsing is adiabatic with respect to the magnetic field gradient induced by the coupling to the nuclei. In this region of detuning, the relevant eigenstates are the antiparallel spin states $(|\uparrow, \downarrow \rangle, |\downarrow, \uparrow \rangle)$. Only the lowest energy state of the two is connected to the singlet state and is created by the pulsing procedure. It is possible to turn on exchange interaction by pulsing the detuning at $\epsilon _0$ for a short and controlled time and induce fast nanosecond coherent oscillations between antiparallel spin states. In this region, the spin qubit is highly sensitive to the electrical environment. The system is then brought back to the measurement position 1 of the (0,2) charge configuration in a mirror procedure. The two Zeeman antiparallel spin states give then two different charge outputs: either S(0,2) or $T_0$(1,1). Altogether, the complete manipulation sequence lasts 12 $\mu s$. Then, the system is kept in the measurement position 1 for a time of 10 $\mu s$. The relaxation time in this position has been measured to be equal to 30 $\mu s$. Since most of the sequence is spent in the measurement position, the conductance measured with a slow IV converter will reflect the charge configuration in this position. Singlet probability is deduced from the level of conductance extracted from the Fig. S2(a).

It is worth noticing that an important requirement to observe coherent oscillations is the fact that the singlet is the ground state in a two-electron quantum dot. In perpendicular magnetic field, this means that the field cannot exceed 1 T \cite{meunier}. This is the reason why our experiment is performed at a filling factor of 16 and not lower.

\begin{figure}
\includegraphics[width=3in]{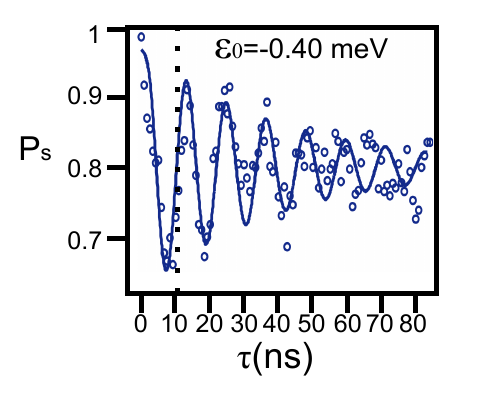}
\caption{\textbf{Reference coherent exchange oscillation at $\epsilon _0 = -0.40$ meV.} The probability $P_s$ is plotted as a function of the pulse duration $\tau$. $V_{bias}$ is zero for this reference curve. The nanosecond detection pulse sequence is decomposed in two parts : we add to a 10.6 ns exchange pulse at  $\epsilon _0 = -0.40$ meV (dashed line) the nanosecond pulse at $\epsilon _d = -0.10$ meV.}
\label{FigS3}
\end{figure}

To analyze quantitatively the coherent oscillations presented in Fig. 2 and Fig. \ref{FigS3}, we restrict the Hilbert space to the antiparallel spin states, we use the $J$ extracted from the 2D plot presented in Fig. 1(b) and we fit numerically the data according to the expected time evolution from the Hamiltonian $H=J(\epsilon(\tau))/2 \sigma _z+ dB_z /2 \sigma _x$ where $dB_z$ is the expected magnetic field gradient due to the coupling to the nuclei of the structure \cite{petta}. To obtain good agreement, we had to introduce an amplitude factor $a$ and the finite rise time $\tau_{RC}$ of the pulse detuning excitation $\epsilon(\tau)$. Both parameters $dB_z$ and $\tau_{RC}$ are considered to be the same and fixed for all curves in the fitting procedure. The decay of the coherent oscillations are fitted with an exponential decoherence law with a characteristic rate $\Gamma$. The output of the fitting procedure yields then three parameters ($a$, $\epsilon_0$ and $\Gamma$). 

In Fig. \ref{FigS3}, we present the coherent oscillations serving as a reference to produce the nanosecond pulse detection.

\textbf{Electron injection in the edge state}

\begin{figure}
\includegraphics[width=3.4in]{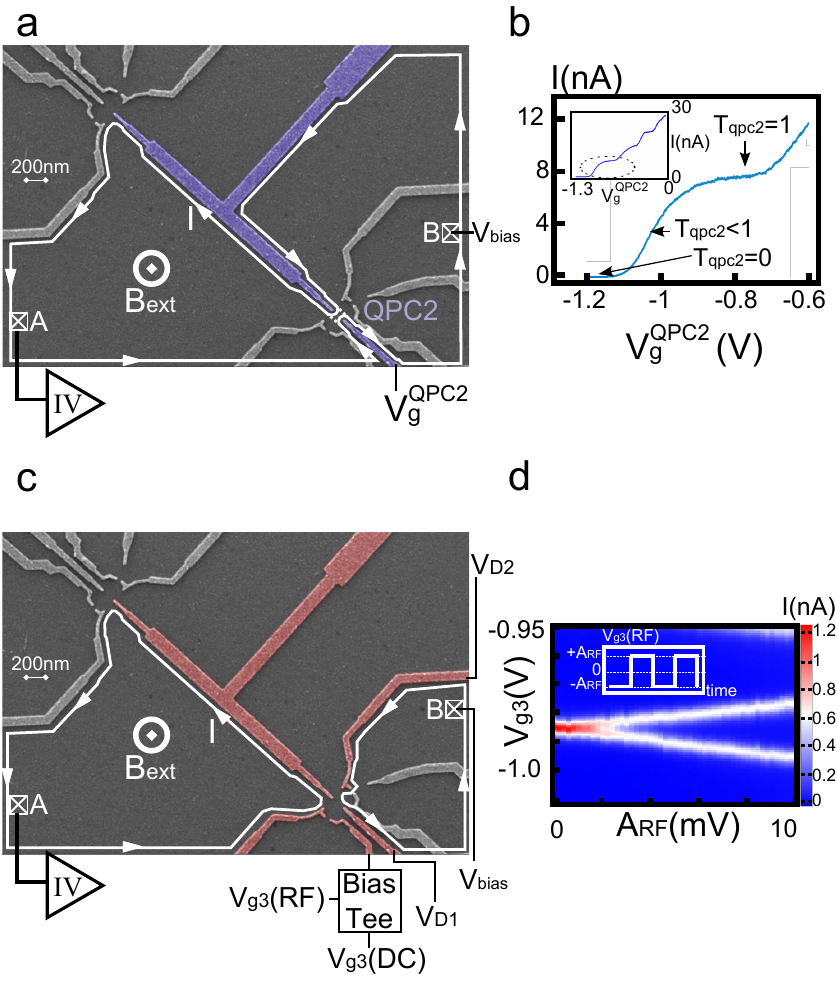}
\caption{\textbf{Injection of electrons in the edge in the QPC. (a)} Scanning Electron Microscope image of the device and schematic of the experimental setup in the QPC configuration. The polarized gates to obtain the QPC are colored in blue. We represent with a white plane line the outer edge channels. \textbf{(b)} Current as a function of the applied voltages on the gate $V_{g}^{QPC2}$. The last conductance plateau is observed demonstrating that we inject the electrons in the closest edge channel from the gates.  \textbf{(c)} Scanning Electron Microscope image of the device and schematic of the experimental setup in the quantum dot configuration. The polarized gates to obtain the quantum dot are colored in red. \textbf{(d)} Up : Current as a function of the DC applied voltages on the fast quantum dot gate $V_{g3}$. Down : a square AC voltage sequence with an amplitude $A_{RF}$ and a repetition rate of 160 ns is added.}
\label{FigS4}
\end{figure}

To inject electrons into the edge state, we use two different protocols. We use a QPC (see Fig. \ref{FigS4}(a)) to inject a stream of electrons. We can tune it on the last plateau where only the last edge state is transmitted (Fig. \ref{FigS4}(b)). By varying the voltages applied on $V_{g}^{QPC2}$, we can vary the channel transmission continuously. To inject on-demand a single electron in a controllable manner into the edge channel, a quantum dot is formed by applying negative voltages on the red colored gates (see Fig. \ref{FigS4}(c)). Transport through the quantum dot can be observed between the contacts A and B. We tune the barriers of the quantum dot in order to achieve the symmetric sequential tunneling. In this case, a current of 1 nA was running through the quantum dot (Fig. \ref{FigS4}(d)) and it results that the tunneling through the barriers was subnanosecond. In order to ensure that the dot is maximally coupled to the reservoir connecting it to the $S-T_0$ detector, we close the barrier connecting the dot to the other reservoir by applying more negative voltages on the two gates $V_{D1}$ and $V_{D2}$. As a result, the current at the top of the Coulomb peak is decreasing by one order of magnitude. We can assume that the dot is only coupled to the right reservoir and that the typical timescale is at least 100 ps. In Fig. \ref{FigS4}(d), we present the resulting current when the dot is excited with a square wave on the gate $V_{g3}$.